\documentclass[aapm,reprint,graphics]{revtex4-1}

\usepackage{graphicx}
\usepackage{amssymb}
\usepackage{float}
\usepackage{color,soul}
\usepackage[mathlines]{lineno} 
\modulolinenumbers[5] 

\begin{document}

\title{Automated target tracking in kilovoltage images using dynamic templates of fiducial marker clusters}

\author{Warren G. Campbell}
\email[e-mail correspondence: ]{warren.campbell@ucdenver.edu}
\author{Moyed Miften}
\author{Bernard L. Jones}
\affiliation{Department of Radiation Oncology, University of Colorado School of Medicine, Aurora, Colorado 80045}

\date{\today}

\begin{abstract}

\textbf{Purpose:}   Implanted fiducial markers are often used in radiotherapy to facilitate accurate visualization and localization of tumors.  Typically, such markers are used to aid daily patient positioning and to verify the target's position during treatment.  These markers can also provide a wealth of information regarding tumor motion, yet determining their accurate position in thousands of images is often prohibitive.  This work introduces a novel, automated method for identifying fiducial markers in planar x-ray imaging.

\textbf{Methods:}  In brief, the method was performed as follows.  First, using processed CBCT projection images, an automated routine of reconstruction, forward-projection, tracking, and stabilization generated static templates of the marker cluster at arbitrary viewing angles.  Breathing data were then incorporated into the same routine, resulting in dynamic templates dependent on both viewing angle and breathing motion.  Finally, marker clusters were tracked using normalized cross correlations between templates (either static or dynamic) and CBCT projection images.  To quantify the accuracy of the technique, a phantom study was performed and markers were manually tracked by two users to compare the automated technique against human measurements.  Then, 75 pre-treatment CBCT scans of 15 pancreatic cancer patients were analyzed to test the automated technique under real life conditions, including several challenging scenarios for tracking fiducial markers (e.g., extraneous metallic objects, field-of-view limitations, and marker migration).

\textbf{Results:}  In phantom and patient studies, for both static and dynamic templates, the method automatically tracked visible marker clusters in 100\% of projection images.  For scans in which a phantom exhibited 0D, 1D, and 3D motion, the automated technique showed median errors of 39 $\mu$m, 53 $\mu$m, and 93 $\mu$m, respectively.  Human precision was worse in comparison; median inter-observer differences for single markers and for the averaged coordinates of 4 markers were 183 $\mu$m and 120 $\mu$m, respectively.  In patient scans, the method was robust against a number of confounding factors.  Automated tracking was performed accurately despite the presence of radio-opaque, non-marker objects (e.g., metallic stents, surgical clips) in 5 patients.  This success was attributed to the distinct appearance of clusters as a whole compared to individual markers.  Dynamic templates produced higher cross-correlation scores than static templates in patients whose fiducial marker clusters exhibited considerable deformation or rotation during the breathing cycle.  For other patients, no significant difference was seen between dynamic and static templates.  Additionally, transient differences in the cross-correlation score identified instances where markers disappeared from view.

\textbf{Conclusions:}  A novel, automated method for producing dynamic templates of fiducial marker clusters has been developed.  Production of these templates automatically provides measurements of tumor motion that occurred during the CBCT scan that was used to produce them.  Additionally, using these templates with intra-fractional images could potentially allow for more robust real-time target tracking in radiotherapy.

\textit{This manuscript was submitted to Medical Physics}

\end{abstract}

\pacs{} 
\maketitle 

\section{Introduction}
\label{Intro}

Tumors are often difficult to identify in photon-based imaging due to similarities between normal and tumor tissue.  A common approach to improve target visibility is to implant radio-opaque fiducial markers in or near the tumor.  These dense, metal objects serve as easy-to-find landmarks of the tumor position in planar kV/MV imaging, or in cone-beam computed tomography (CBCT).  In radiation therapy, a physician will examine these markers in a CBCT scan acquired just prior to treatment in order to localize the tumor with the isocenter of the treatment beam.\cite{Jaffray:2002tj, Moseley:2007dc}  In some instances, intra-fractional monitoring of these markers can also allow for positional verification of the target during treatment, often to accommodate respiratory gating.\cite{Seppenwoolde:2002um, Berbeco:2005fd}  Furthermore, offline review of CBCT scans can provide valuable data on the motion of fiducial markers, which can inform choices related to margin selection and motion management.\cite{Poulsen:2008hs}  The latter two cases call for automated techniques rather than manual techniques, due to a need for quick reaction time and the high workload involved, respectively.

Many methods for locating radio-opaque markers in x-ray images are based on template matching.\cite{Lam:1993uf, Lewis:1995ue, Balter:1995hy, Harris:2006bg, Tang:2007jv, Cho:2009bj, Poulsen:2011dq, Fledelius:2011gb, Regmi:2014eb, Fledelius:2014gq, Jones:2015dj}  Simply, template matching attempts to find an object in a sample image using a template image that is representative of that object.  A 2D normalized cross-correlation of the template image and a sample image provides a pixel map of cross-correlation scores ranging between -1 and 1.\cite{Lewis:1995ue}  A pixel with a score of 1 indicates that the exact object is located at the pixel's location, and a score of -1 is indicative of the exact inverse of the object (i.e., a `negative' image of the object).  Essentially, template matching identifies high-contrast regions of an image whose shape resembles that of the template.  When trying to detect fiducial markers used for radiation therapy, the process can be obscured by other high-contrast features of the image, such as bony anatomy, air pockets, or other metallic objects.

Marker templates are typically based on the properties of a single marker.  Templates can be prepared for spherical and cylindrical markers by knowing their dimensions and the geometry of the imaging setup.\cite{Lam:1993uf, Fledelius:2011gb, Fledelius:2014gq}  For cylindrical markers, imaging ahead of time can help to determine their position and orientation, allowing for templates to be prepared according to imaging angle.  Some markers require that images be taken in advance due to their arbitrary shape and deformation that can occur during implantation (e.g., coil markers).\cite{Poulsen:2011dq, Regmi:2014eb}  These so called `coaching' images can be used to produce templates for subsequent tracking.  However, a cyclical challenge presents itself whenever such coaching images need to be used for a template production technique that is meant to be fully automated.  That is, how does one reliably and in a fully automated manner detect markers in coaching images in order to create templates that will later be used to reliably and in a fully automated manner detect the same markers in clinical images?

Thus far, methods for template generation have either required some form of manual selection by the user, or required that assumptions be made about the shape of markers.  In recent years, Poulsen et al reconstructed templates for arbitrarily shaped markers from CBCT projection images, but these images had to be selected manually and required uniform backgrounds.\cite{Poulsen:2011dq}  A method by Fledelius et al assumes that markers are cylindrical in shape, and relies on building a constellation model for each individual marker.\cite{Fledelius:2011gb, Fledelius:2014gq}  Regmi et al introduced a semi-automated technique for arbitrarily shaped markers; however, their technique built templates using planning CT data with relatively poor resolution, required the manual selection of a volume of interest from the planning CT to locate markers, and required manual corrections for variations in inter-marker spacing.\cite{Regmi:2014eb}

In this work, we introduce a novel method of template image generation that uses a single CBCT scan.  This method is fully automated, requires no input from the user, and makes no assumptions about the shapes of markers.  Instead of searching for fiducial markers individually, this method treats the cluster of markers as a single entity, seeking out the cluster as a whole.  The entire cluster is less likely than individual markers to be confused with other aspects of a patient's anatomy.  Also, searching for the cluster as a whole eliminates the challenge of needing to differentiate individual markers when they overlap with one another.

\begin{figure*}
\includegraphics[width=17.75cm]{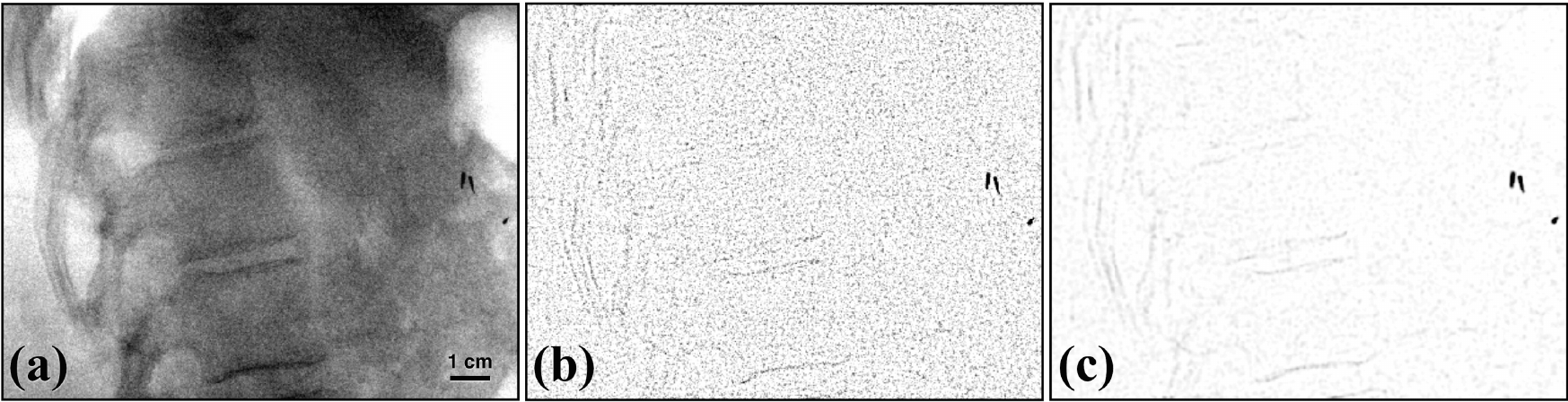}
\caption{\label{FigA}  Three fiducial markers can be seen in these examples from patient \#8 of (a) a projection image, (b) a marker-enhanced image, and (c) a filtered marker-enhanced image.  The scale indicator in (a) shows 1 cm as projected at the isocenter.}
\end{figure*}

\section{Materials \& Methods}
\label{MatsMeths}

\subsection{Data Acquisition}
\label{MMData}

\subsubsection{Cone-Beam Projection Data}
\label{CBCTdata}

Cone-beam CT scans were acquired using the on-board imager of a TrueBeam\texttrademark$ $ STx linear accelerator (Varian Medical Systems; Palo Alto, CA).  In each scan, 892 projection images (768$\times$1024, 125 kV, 80 mAs) were collected over a full rotation using a half-fan geometry with square pixels 0.388 mm in size (0.259 mm projected at the isocenter).  Images were acquired at a rate of 14.8 images per second.  For patient scans (discussed further in Section \ref{PatientData} below), the Real-Time Position Management\texttrademark$ $ (RPM) system (Varian Medical Systems) was used to collect breathing data.  Projection images and pertinent metadata were extracted from CBCT scan files using MATLAB (MathWorks; Natick, MA) and specialized scripts provided by Varian.

\subsubsection{Phantom Data}
\label{PhantomData}

Scans of a heterogeneous thorax phantom (CIRS, Inc.; Norfolk, VA) were acquired to evaluate the accuracy of the template tracking technique.  Four fiducial markers (gold, cylindrical, 5 mm length, 1 mm diameter) were positioned inside of the phantom, and movement of the phantom was performed using three orthogonal, linear robotic stages (Velmex, Inc.; Bloomfield, NY).  Three scans of the phantom were acquired: one where no motion was imposed (0D), one where motion was imposed only in the superior-inferior direction (1D), and one where motion was imposed in all three directions (3D).  The magnitude of motion imposed on the phantom was chosen to be representative of typical motion observed in the pancreatic cancer patients also evaluated in this work (left-right, anterior-posterior, and superior-inferior ranges of motion of 5 mm, 5 mm, and 10 mm, respectively).\cite{Jones:2015jz}

\subsubsection{Patient Data}
\label{PatientData}

Seventy-five CBCT scans were acquired of 15 patients receiving stereotactic body radiation therapy for pancreatic cancer.  These were routine patient alignment scans taken just prior to the delivery of each of their five treatment fractions.  Prior to simulation and treatment planning, each patient had 3 to 4 fiducial markers (titanium-coated carbon, roughly cylindrical, 5 mm length, 1 mm diameter) implanted in their tumor in order to aid daily 3D target localization.  For the purposes of motion mitigation, abdominal compression was used for all 15 patients.  To allow for the collection of breathing data during each scan, the RPM external marker block was positioned on the patient's upper abdomen, roughly midway between the superior edge of the compression belt and the patient's xiphoid process.

\subsection{Template Production}
\label{Templates}

\begin{figure*}
\includegraphics[width=17cm]{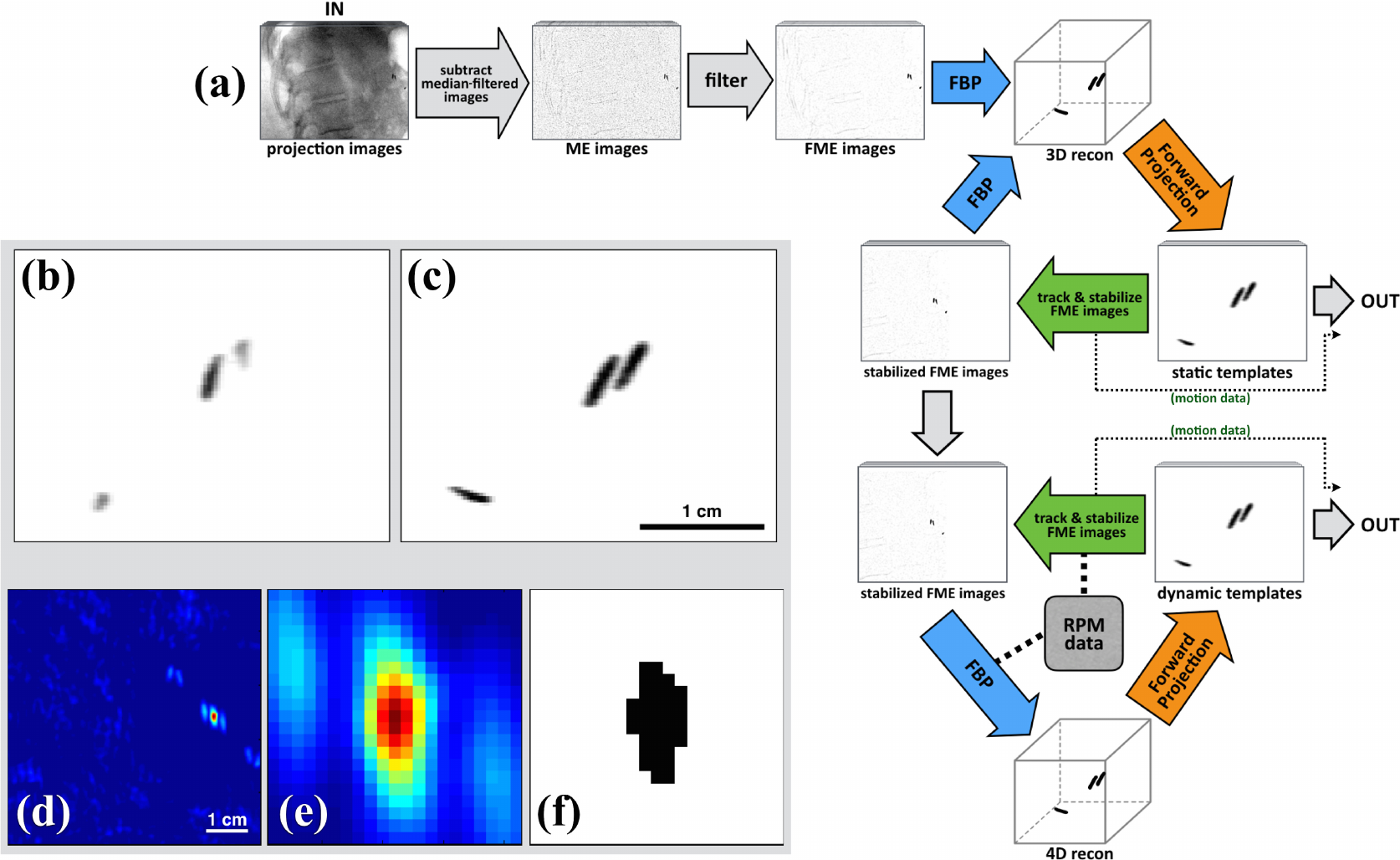}
\caption{\label{FigB}  Production of static and dynamic templates: (a) flowchart of the entire process, (b) example of a template produced from an initial, crude reconstruction, and (c) the template produced after 3 iterations of the track \& stabilize, reconstruct, and forward-project loop.  Using completed dynamic templates, examples from patient \#8 are shown for: (d) a normalized cross-correlation of a template with an FME image, (e) the local window, and (f) the mask of FAHM pixels.  The scale indicators in (c) and (d) show 1 cm as projected at the isocenter, and both (e) and (f) are square with sides equaling 5.43 mm.}
\end{figure*}

Broadly, our method is outlined as follows.  First, CBCT projection images are processed to enhance the appearance of markers (see Figure \ref{FigA}).  Then, a set of static templates (dependent on gantry angle) is created from these processed images through an iterative routine that uses filtered back-projection to reconstruct the cluster, forward-projection to create template images, and template tracking to correct for motion seen in projection images.  Once this loop converges on a set of static templates, breathing data is incorporated into the same process to construct a set of 4D dynamic templates (dependent on both gantry angle and breathing motion).  To aid the more detailed description that follows, Figure \ref{FigB} provides a flowchart illustrating the entire template production process.

To enhance the appearance of fiducial markers, which have an intensity value lower than surrounding pixels, CBCT projection images were processed using filtering methods in MATLAB.  First, median-filtered versions of each projection image (\emph{medfilt2}, 9$\times$9) were calculated.  Then, these median-filtered images were subtracted from their respective originals, resulting in marker-enhanced (ME) images.  In ME images, pixels with values greater than zero were set equal to zero, leaving behind the highly radio-opaque markers, some edge features (e.g., from bones and gas in the bowels), and some random noise.  Next, ME images were filtered in sinogram space in all three directions using a smoothing filter (\emph{sgolayfilt}, k=3, f=5).\cite{Savitzky:1964bn}  Then, the values of each individual pixel were examined across the entire scan, and for any projections during the scan when a pixel's value rose above its 40\textsuperscript{th} percentile value, its value for that projection was set equal to zero.  This step is based on the assumption that markers are not likely to remain exactly at the isocenter, so markers should pass across any given pixel for only a portion of the scan (i.e., no more than 40\% of the scan's duration).  Finally, each projection image was filtered again using a median filter (\emph{medfilt2}, 3$\times$3) and then an adaptive noise-removal filter (\emph{wiener2}, 3$\times$3).\cite{Wiener:1949}  The resultant filtered marker-enhanced (FME) images were used for template production and motion tracking.  Figure \ref{FigA} shows examples of a projection image and its respective ME and FME images.

Filtered back-projection was used with FME images to provide a crude, initial reconstruction of the fiducial marker cluster.  In the reconstructed volume, voxels with values below a threshold equal to 70\% of the maximum value were set equal to zero and Gaussian filtering was used to clean up the resultant volume.  The non-zero values that remained were assumed to represent the marker cluster, and these values were shifted so that their 3D center of mass aligned with the center of the volume.  Then, forward-projection was used to produce 360 template images as a function of gantry angle (see Figure \ref{FigB}b).  Using this set of template images, the position of the cluster was tracked throughout the scan using normalized cross-correlations of template images and FME images (tracking is discussed in more detail in Section \ref{Tracking}).  Once cluster positions were obtained, each FME image was stabilized by centering the cluster in frame.  Using filtered back-projection with stabilized FME images, the volume that is reconstructed suffers from fewer motion artifacts and better resembles the fiducial marker cluster.  By repeating this loop of (i) tracking \& stabilizing FME images, (ii) reconstructing stabilized data, and (iii) forward-projecting template images, a set of high quality static templates---360 images as a function of gantry angle---could be obtained (see Figure \ref{FigB}c).  Typically, high quality templates were produced after 3 iterations, but the loop continued until no significant increase in cross-correlation scores was observed.

After static templates were prepared, breathing data could be incorporated into the same iterative routine to produce dynamic templates that are a function of both gantry angle and RPM motion.  Using stabilized FME images, 5 different volumes were reconstructed for 5 different ranges of RPM displacement.  From each of these 5 volumes, a set of template images based on gantry angle were produced.  In order to limit sudden changes in templates during tracking, the 5 ranges of breathing amplitude used for reconstruction were deliberately chosen to include overlap.  Without overlapping ranges, a single template would be used for a given range of RPM positions, and templates would suddenly change when the RPM's position transitioned from one range into another.  Instead, with overlapping ranges, a weighted average of two overlapping templates was used.  In this fashion, templates will not exhibit sudden changes when the position of the RPM transitions between ranges.  Based on patient observations, which indicate longer dwell times in the end-exhalation phase, ranges for the 5 breathing amplitude bins were selected to include the following percentiles: 0-40, 25-55, 40-70, 55-85, and 70-100, where 0 indicates end-exhale and 100 indicates end-inhale.  With the exception of values below 25 and above 85, templates selected during tracking were always a weighted-average of two templates, with weightings being dependent on the distance of the current RPM position from the center of the two encompassing ranges.  After the loop of (i) tracking \& stabilizing, (ii) reconstructing stabilized data, and (iii) forward-projecting template images was repeated 2 or 3 times, a set of high quality dynamic templates---360$\times$5 images as a function of gantry angle and RPM position, respectively---were produced.

\subsection{Marker Tracking}
\label{Tracking}

\subsubsection{Automated Tracking}
\label{AutoTrack}

Tracking with template matching determines the position of the target (i.e., the marker cluster) by locating the correct peak in the normalized cross-correlation of the projection image and the template.  The correct peak, however, is not always the peak with the global maximum.  Occasionally during a CBCT scan, image features that are not the true target (e.g., patient anatomy, foreign objects) can resemble the target, resulting in an erroneous peak.  Almost always, these resemblances are short-lived, only bearing a similarity at certain angles.  As such, it is common to only search for peaks in a local window, reducing the likelihood of selecting an erroneous peak.  This window can be chosen based on the expected position of the marker as predicted by a planning CT.  Although, a fully automated tracking technique would preferably not depend on this relatively old and possibly unreliable data.

For the automated technique in this work, the global maximum was initially located in each cross-correlated image for the entire scan.  Next, it was assumed that the longest consecutive chain of global peak positions (i.e., without large displacements between frames) corresponded to the correct peak.  Then, from the midpoint in this chain, forward tracking and backward tracking (i.e., towards the first and last images in the scan, respectively) was performed by locating the local peak within a window centered on the previously tracked position.  To allow for sub-pixel precision, the position of each peak was calculated as the center of mass of the 9 pixels within the local tracking window with the highest cross-correlation scores, regardless of whether or not these pixels were adjacent to one another.

Figure \ref{FigB}d shows an example of a normalized cross-correlation (\emph{normxcorr2}) of a template with an FME image.  For this work, a local window of 21$\times$21 pixels (5.43$\times$5.43 mm$^{2}$ projected at the isocenter) was used for tracking (see Figure \ref{FigB}e).  Three factors informed the choice of dimensions for this window: (i) pixel dimensions, (ii) imaging frequency, and (iii) maximum speeds of pancreatic tumors observed during a previous work.  For an accumulated set of 97 pancreatic cancer patients, the maximum instantaneous tumor speed observed was 3 cm/s.\cite{Jones:2015jz}  At this speed, with pixels being 0.259 mm projected at the isocenter and images being acquired every 0.0676 seconds, one could expect the position of a cluster to be displaced up to 8 pixels between frames.  A 17$\times$17 window would be capable of catching such displacements if only single pixel maxima were being used to localize the cluster.  However, in this work, the centroid of the 9 pixels with the highest values within the local window were used to calculate the cluster's position.  Thus, in order to allow for the consideration of a cluster of 9 pixels, the local window was expanded to fully encompass 2 pixels beyond the maximum displacement.

To reduce computing time, only a region-of-interest (ROI) one quarter of the size of the imaging panel (384$\times$512) was ever considered in this work.  This ROI was centered about the isocenter in the superior-inferior direction.  Due to the half-fan geometry of each scan, this ROI was flush with the edge of the imager in the lateral direction so that as much of the patient near the isocenter could be observed.  Cropping of the imaging panel in this manner still allowed for entire fiducial marker clusters to be fully visible for all scans observed in this work, with the exception of incidences when clusters fell outside the field-of-view of the imaging panel due to the half-fan lateral shift.

\subsubsection{Manual Tracking}
\label{ManualTrack}

In addition to automated tracking for all scans, two users manually tracked the positions of fiducial markers in the 3 phantom scans, and one user manually tracked markers for 3 scans from 3 patients.  These measurements were partly assisted by measurements obtained by the automated tracking technique, in that automatically tracked positions were used to show the user a zoomed-in region of each projection image, automatically magnifying the cluster.  The MATLAB function \emph{ginput} was used to convert mouse clicks into sub-pixel measurements of the center of each marker, as judged by the user.  One user repeated these measurements so that both inter-observer and intra-observer precision could be evaluated.

\begin{figure*}
\includegraphics[width=17.75cm]{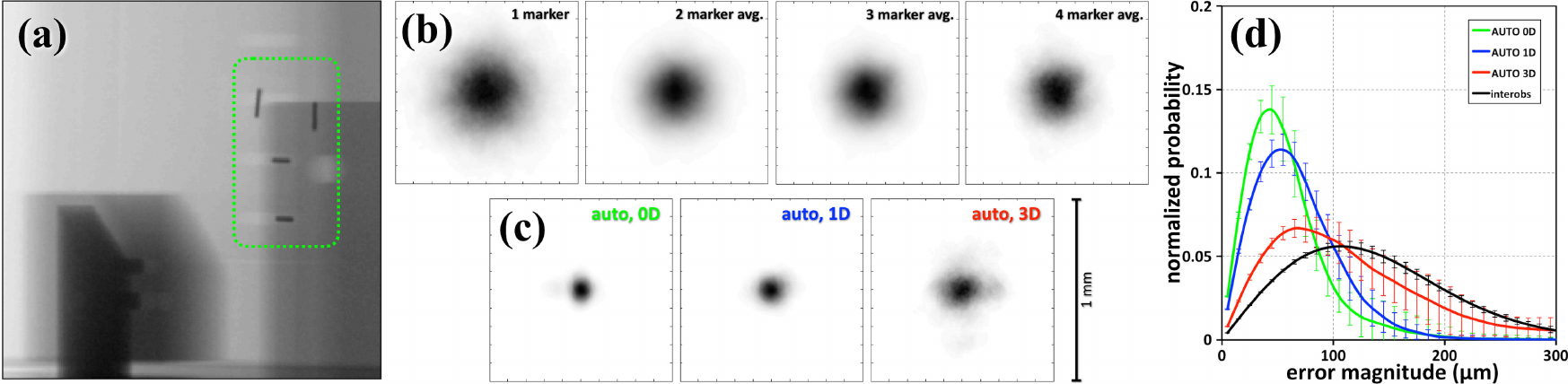}
\caption{\label{FigG}  Results from the phantom study: (a) projection image with 4 markers indicated, (b) distributions of inter-observer differences for manually selected positions of individual markers and for the averaged coordinates of up to 4 markers, (c) distributions of errors for the automated tracking technique in the 0D, 1D and 3D scans, and (d) normalized probability distribution functions for errors in the 0D, 1D, and 3D scans, and for inter-observer differences of the 4-marker average.  All sides in (b) and (c) are 1 mm in length.  Error bars in (d) are equal to one standard deviation.}
\end{figure*}

\subsection{Evaluation}
\label{Eval}

In the phantom study, inter-observer and intra-observer differences of manually determined marker positions were calculated for each marker in each projection image.  Furthermore, the inter-observer and intra-observer differences of the average position of multiple markers were calculated, up to and including all four markers.  Due to the fact that the automated tracking technique seeks the entire cluster as a whole, it is only fair to compare automatically tracked positions against the averaged user-determined positions of all four markers.  Known positions were calculated by smoothing the averaged measurements of all four markers and both users.  In this way, sag of the imager was also accounted for.  Errors in the phantom study were calculated as the magnitude of the 2D difference between the automatically tracked position and the known position.  The criterion for successfully accurate tracking in the phantom study was an error \textless1 mm (projected at the isocenter).

For patient data, in addition to tracking the locations of marker clusters, 3 cross-correlation metrics were recorded during automated tracking to evaluate the strength of template matches.  The first metric recorded during tracking was the maximum score of any single pixel within the local window, with scores closer to 1 indicating a better match.  The second metric recorded was the global maximum score observed within the entire quarter-sized ROI area of the imaging panel.  With this metric, we could monitor the global tracking rate---the rate at which the local maximum was also the global maximum.  Whenever the local maximum dropped below the global maximum, we knew that our ability to continuously track the cluster was dependent on the use of a local search window.  Finally, the third metric recorded was the number of pixels in the local window with values equal to or greater than half of the maximum value.  By multiplying these numbers by the area of a single pixel as projected at the isocenter, we effectively monitor the full-area-half-max (FAHM) of the cross-correlation peak (see mask in Figure \ref{FigB}f).  Akin to the full-width-half-max for point spread and line spread functions, a smaller FAHM would denote a sharper peak, indicating a more precise match.  Using manually tracked data from three patient scans, a relationship between FAHM and error magnitude was established, where error magnitude was calculated as the difference between automatically tracked positions and manually selected positions.  With this relationship, inferred absolute error values were calculated for all patient data in order to provide an indication of accuracy in real-life clinical scenarios where the ground truth is unknown.  For patient data, the accuracy of tracked positions was verified by a visual check of stabilized scans, looking for any noticeable displacements during each scan.

\section{Results}
\label{Results}

\subsection{Phantom Study}
\label{PhantomStudy}

For all analysis in the phantom study, 2D differences and 2D errors were calculated relative to the coordinates of the imaging panel in units of micrometers as projected at the isocenter.  Figure \ref{FigG}b shows distributions of inter-observer differences for single markers and for the averaged position of 2, 3, and 4 markers.  The median magnitude of these differences were 183 $\mu$m, 145 $\mu$m, 129 $\mu$m, and 120 $\mu$m, respectively.  For intra-observer differences, these values were 149 $\mu$m, 120 $\mu$m, 106 $\mu$m, and 98 $\mu$m, respectively.

Automated tracking in the phantom study was performed using static templates instead of dynamic templates due to the fact that no deformation or rotation of the marker cluster was possible with our experimental setup.  Figure \ref{FigG}c shows error distributions for the automated tracking technique in the 0D, 1D, and 3D phantom scans.  Median error magnitudes for these scans were 39 $\mu$m, 53 $\mu$m, and 93 $\mu$m, respectively.  Errors in all three scans were less than 1 mm (i.e., were successfully accurate) and had 99\textsuperscript{th} percentiles of 152 $\mu$m, 174 $\mu$m, and 324 $\mu$m, respectively.  Part of the increase in error magnitudes seen in the 1D and 3D scans was attributed to unavoidable vibrations of the phantom stage that occurred during motion.  Normalized probability distributions for the automated tracking technique in the 0D, 1D, and 3D scans are shown in Figure \ref{FigG}d, along with the distribution for inter-observer differences of the averaged position of 4 markers.

\begin{figure}[t]
\includegraphics[width=7.5cm]{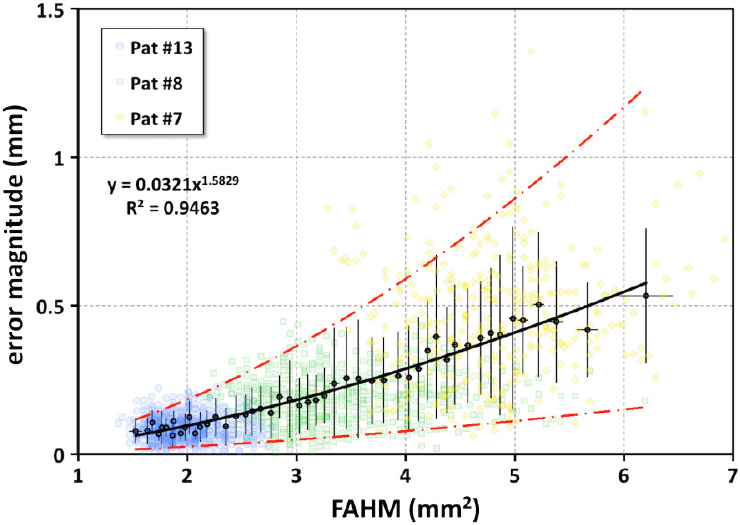}
\caption{\label{Fig4}  Manually selected marker positions for three patients were used to establish a relationship between full-area-half-max (FAHM) and error magnitude.  Using this relationship, FAHM was used to calculate inferred absolute errors (IAE) for patient data.  Error bars shown are equal to one standard deviation, and broken red lines indicate 5\textsuperscript{th} and 95\textsuperscript{th} percentiles.}
\end{figure}

\begin{table*}
\begin{tabular}{ccccc}\toprule
{  }{  \textbf{patient \#}  }{  } & {  }{  \textbf{\# of markers}  }{  } & {  }{  \textbf{xcorr score}  ({\boldmath$\mu$} $\pm$ $\sigma$)  }{  } & {  }{  \textbf{global tracking rate}  [\%]  }{  } & {  }{  \textbf{IAE}  (\boldmath$\mu$, 90\% range) [$\mu$m]  }{  } \\
\colrule
1 & 4 & \textbf{0.79} $\pm$ 0.09 & 99.8 & \textbf{214}, 128--283 \\
2 & 3 & \textbf{0.86} $\pm$ 0.07 & 100 & \textbf{140}, 44--206 \\
3 & 2 & \textbf{0.78} $\pm$ 0.10 & 88.3 & \textbf{140}, 31--210 \\
4* & 3 & \textbf{0.71} $\pm$ 0.09 & 96.5 & \textbf{158}, 31--206 \\
5 & 4 & \textbf{0.84} $\pm$ 0.05 & 100 & \textbf{181}, 89--193 \\
6 & 3 & \textbf{0.80} $\pm$ 0.10 & 99.9 & \textbf{173}, 89--268 \\
7 & 4 & \textbf{0.48} $\pm$ 0.07 & 96.4 & \textbf{371}, 268--459 \\
8 & 3 & \textbf{0.88} $\pm$ 0.04 & 100 & \textbf{255}, 128--369 \\
9* & 2  /  1& \textbf{0.51} $\pm$ 0.10  /  \textbf{0.82} $\pm$ 0.15&66.0  /  77.0& \textbf{118}, 55--184  /  \textbf{165}, 102--231 \\
10* & 3 & \textbf{0.76} $\pm$ 0.10 & 99.9 & \textbf{189}, 102--274 \\
11 & 3 & \textbf{0.79} $\pm$ 0.12 & 98.9 & \textbf{198}, 136--259 \\
12* & 4 & \textbf{0.57} $\pm$ 0.16 & 91.6 & \textbf{220}, 65--313 \\
13 & 4 & \textbf{0.83} $\pm$ 0.05 & 100 & \textbf{102}, 65--136 \\
14* & 3 & \textbf{0.60} $\pm$ 0.12 & 87.9 & \textbf{102}, 44--170 \\
15 & 3 / 3 & \textbf{0.91} $\pm$ 0.03 / \textbf{0.87} $\pm$ 0.05 & 100 / 99.1 & \textbf{193}, 55--306 / \textbf{128}, 44--200 \\
\botrule
\end{tabular}
\caption{\label{Rtable} Summary of metrics obtained for each patient using dynamic templates.  Note: patients \#9 and \#15 had migration events (see Figure \ref{FigF}); as such, results are given as pre- / post-migration.  *Patients having radio-opaque, non-marker objects in the imaging plane.}
\end{table*}

\subsection{Patient Studies}
\label{PatientStudy}

Marker positions were manually tracked in three CBCT scans from three patients whose FAHM values span the range of values seen in this set of patients.  Figure \ref{Fig4} illustrates the relationship observed between FAHM and error magnitude for all three patients.  With this relationship, FAHM values obtained during the automated tracking of patient data were used to calculate inferred absolute error (IAE) values.

For all patients, the automated technique tracked marker clusters accurately in 100\% of projections images wherein the entire cluster was visible, regardless of whether the templates used were static or dynamic.  Beyond this tracking rate, Table \ref{Rtable} provides a patient-by-patient summary of 4 other noteworthy variables: (1) the number of markers in the cluster, (2) local maximum cross-correlation scores obtained during tracking, (3) global tracking rates, and (4) inferred absolute error values as calculated by FAHM values.  All values shown in Table \ref{Rtable} were based on tracking that used dynamic templates, as they provided higher cross-correlation scores than static templates for select patients (discussed further in Section \ref{RsVd}).  Also, only projection images in which the entire marker cluster was visible in the field-of-view were considered for the data shown in Table \ref{Rtable}.

Tracking metrics varied from patient-to-patient, but overall ($\mu\pm\sigma$) cross-correlation scores were 0.76 $\pm$ 0.12, global tracking rates were 94.9\% $\pm$ 7.5\%, and inferred absolute errors ($\mu$, 90\% range) were 179 $\mu$m, 87--251 $\mu$m. Particularly noteworthy is the global tracking rate, because it provides an indication of the uniqueness of a cluster's appearance amongst the patient's anatomy and any foreign objects. Figure \ref{FigE}a shows the relationship between global tracking rates with respect to the number of markers in the cluster. As one might expect, increasing the number of markers in a cluster tends to increase the global tracking rate.

\subsubsection{Cases with Radio-Opaque, Non-Marker Objects}
\label{RnonMark}

\begin{figure*}
\includegraphics[width=17.75cm]{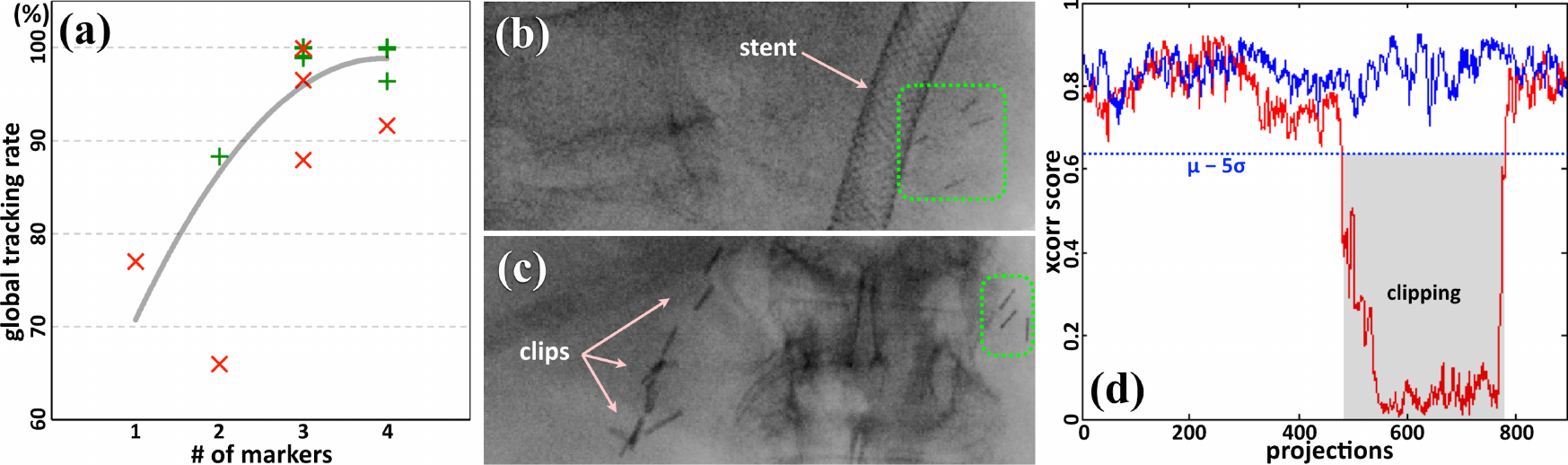}
\caption{\label{FigE}  Global tracking rates are plotted in (a) with respect to the number of markers in the cluster, with data from patients having radio-opaque non-marker objects in the imaging plane shown as red `$\times$'s, and all others as green `$+$'s.  Examples of radio-opaque non-marker objects: (b) a metallic biliary stent from patient \#12, and (c) a large cluster of surgical clips from patient \#10.  In both (b) and (c), dashed green boxes indicate the target fiducial cluster.  In (d), cross-correlation scores for two fractions of the same patient (\#5) are shown, with the cluster falling out of the field of view during one of the fractions; the dashed line indicates 5 standard deviations below the mean for the fraction that did not suffer from clipping.}
\end{figure*}

For 5 patients, medical procedures prior to their radiation treatment resulted in other metallic objects being in the same imaging plane as their fiducial markers.  Such objects can present serious challenges to tracking techniques that look for radio-opaque markers.  In 4 of these patients (\#4, \#9, \#12, and \#14), the non-marker object was a metallic biliary stent used to alleviate a blocked bile duct (see Figure \ref{FigE}b).  In one patient (\#10), a large cluster of surgical clips was visible in the treatment plane (see Figure \ref{FigE}c).  Both of these object types pose unique challenges for any marker tracking technique.  Metallic stents consist of a collapsible wire mesh that, once re-expanded, appears in projection images as a relatively large area of angled lines.  Although, on their own, these stents are not likely to be confused for individual markers, markers can easily be camouflaged by stents whenever they overlap in projection images.  Nevertheless, accurate tracking was maintained for all patients with biliary stents.  The staples seen in Figure \ref{FigE}c present an even bigger challenge for any technique that intends on tracking cylindrical markers.  Despite the fact that these clips are quite similar in appearance to the markers being tracked, automated tracking was still successful, and the global tracking rate was an impressive 99.9\%.

It is also worth noting that automatic template production performed well with both types of radio-opaque, non-marker objects.  For metallic stents, remnants of the wire mesh were apparent in the initial crude templates.  However, the more predominant appearance of the marker cluster meant that subsequent iterations of tracking, stabilizing, and reconstructing end up homing in on the target marker cluster, and motion of the stent mesh relative to the cluster's position caused it to be blurred, falling below the `70\% of maximum density' threshold.  In the case of the cluster of surgical clips, the cluster of clips was not included because it was too far from the isocenter to be considered in the template reconstruction volume.  Hypothetically, if these clips were situated more closely to the target cluster, manual intervention might have been required to ensure that they were not included in the reconstructed cluster.

\subsubsection{Cases with Insufficient Field-of-View}
\label{Rfoview}

In 48 of the 75 CBCT scans examined, one or more markers moved outside the field-of-view during portions of the scan due to the lateral shift of the on-board imager that is imposed by a half-fan CBCT geometry.  In some of these instances, tracking accuracy near the edge of the imager could be maintained by modifying templates based on the last known position of the cluster.  When clusters moved near the edge of the field-of-view, templates were cropped `on-the-fly' based on the proximity of the previously tracked position to the edge of the imaging panel (i.e., portions of the template occurring off of the imaging panel were set equal to zero).  As long as enough of the cluster remained visible---roughly half of the cluster---tracking could still be sustained.  Such cropping allowed for more motion data to be salvaged from these clipped scans.  A visual check of the stabilized scan was performed to verify the validity of tracking measurements in these cases.  In 16 scans, more than half of the cluster fell outside the field-of-view at some point, and tracking of the cluster became either inaccurate or impossible at these positions.  Nevertheless, by continuing to follow peaks at the edge of the imager, tracking was able to automatically resume accurately once the cluster returned into view in 13 of these 16 scans.  In instances where portions of the cluster moved outside of the field-of-view, tracking could be resumed by once again finding the longest consecutive chain of global peaks in the subsequent set of projection images.

Analysis of these events helped to provide some insight into how tracking failures might be recognized when using the current technique.  For these cases, tracking was repeated without implementing the template cropping technique described above.  When clipping occurred, significant drops in cross-correlation scores were observed.  An example of this is shown in Figure \ref{FigE}d, with the clipping event corresponding exceptionally well with the portion of the scan where cross-correlation scores dropped more than 5 standard deviations below the mean score observed in another scan from the same patient in which a clipping event did not occur.  Due to the range of average cross-correlation scores seen in individual patients, it is likely that a single threshold value would not be suitable for catching detection errors for all patients.  Instead, statistically significant drops in cross-correlation scores for each patient would be more appropriate.  If one can determine the point at which a cluster has been lost from the field-of-view, one could continue to monitor the last known position of the cluster instead of following peaks at the edge of the imager, potentially improving the technique's ability to resume tracking after such clipping events.

\subsubsection{Cases with Marker Migration}
\label{RMigrate}

Three patients had noticeable marker migration events.  One patient (\#3) had a marker dislodge prior to their first radiation treatment, resulting in 2 markers remaining in place and the dislodged marker visibly loose in the patient's abdominal cavity.  One patient (\#9) had one marker dislodge before the first treatment fraction, and a second marker dislodge after the first fraction, resulting in only 1 marker remaining in place for the last 4 fractions.  One patient (\#15) had 1 of 3 markers noticeably shift its position relative to other markers after 3 treatment fractions.  Patients \#9 and \#15, having had CBCT scans before and after a migration event, served as useful cases to evaluate how robust the current technique would be against marker migration.  As such, marker clusters were tracked in post-migration CBCT scans using two different sets of dynamic templates: templates produced from images obtained on the treatment day prior to the migration event (pre-migration templates), and templates produced from images obtained on the treatment day after the migration event (post-migration templates).  With the assumption that post-migration templates would more accurately track marker clusters in the post-migration scan, the difference between these two motion measurements would be indicative of the magnitude of errors one could expect if new templates were not produced after a migration event.

\begin{figure}[t]
\includegraphics[width=8.5cm]{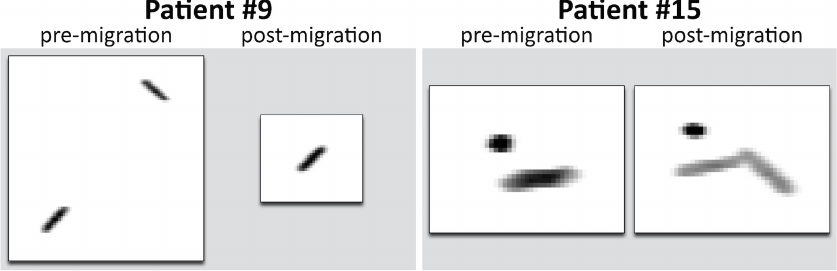}
\caption{\label{FigF}  Two patients had noticeable marker migration events between treatment fractions.  Patient \#9 had a marker dislodge, leaving one marker in place.  Patient \#15 had one marker shift its position relative to the other two.  Pre-/post-migration templates are shown from the same gantry angle.}
\end{figure}

Pre-migration and post-migration templates for patients \#9 and \#15 are shown in Figure \ref{FigF}.  In the case where pre-migration templates were used to track motion in the post-migration scan, tracking was still successfully automated.  After correcting for differences in the centroid positions of pre- and post-migration templates, the maximum difference between positions tracked by pre- and post-migration templates was 203 $\mu$m for patient \#9, and 91 $\mu$m for patient \#15.  Future work will more precisely examine magnitudes of inter-fraction migration seen for individual markers within a cluster.

\subsubsection{Static Templates vs. Dynamic Templates}
\label{RsVd}

\begin{figure*}
\includegraphics[width=17.75cm]{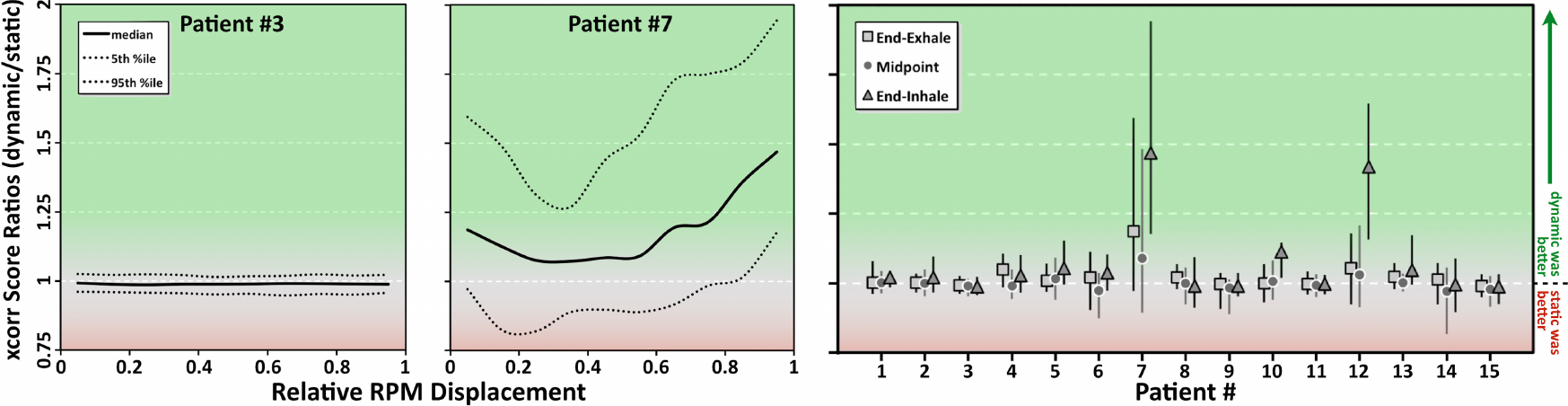}
\caption{\label{FigD}  Ratios of cross-correlation scores obtained using either dynamic or static templates.  In addition to three-point summaries for each individual patient, ratios with respect to relative RPM displacement are shown for patients \#3 and \#7---the patients benefiting the least and the most from the use of dynamic templates, respectively.  A ratio greater than 1 indicates that cross-correlation scores were higher for dynamic templates.  Error bars show 5\textsuperscript{th}--95\textsuperscript{th} percentile ranges of ratio values.}
\end{figure*}

Most patients saw little difference in their cross-correlation scores when dynamic templates were used instead of static templates.  Figure \ref{FigD} provides a summary of cross-correlation score ratios (dynamic/static), with ratios greater than 1 indicating that higher cross-correlation scores were obtained by dynamic templates.  Sample plots of score ratios with respect to relative RPM displacement are shown for patient \#3 and patient \#7, the patients benefiting the least and the most from dynamic templates, respectively.  End-exhalation, midpoint, and end-inhalation points are plotted for all patients.

Patient \#7 and patient \#12 both benefited significantly from dynamic templates, with patient \#7's cluster exhibiting considerable deformation during their breathing cycle, and patient \#12's cluster exhibiting considerable rotation.  For these two patients, the highest benefit when using dynamic templates was seen in end-inhalation phases, which is to be expected.  Static templates represent an average of the cluster's appearance throughout the breathing cycle.  With more time being spent in end-exhalation and mid-range phases, static templates tend to be more representative of the cluster during those intervals.  Dynamic templates allow for images to be prepared that are more representative of the cluster throughout all phases of the breathing cycle.  Although the majority of patients did not exhibit levels of deformation or rotation that necessitated the use of dynamic templates, it should be noted that the use of dynamic templates did not have a detrimental effect on cross-correlation scores for these patients.  As such, in a clinical scenario where breathing data were available, dynamic templates could be used by default for all patients.

\section{Discussion}
\label{Discuss}

The field of radiation therapy is trending towards higher dose per fraction delivery (i.e., stereotactic body radiation therapy, SBRT), as well as increased usage of image guidance (both pre-treatment and intra-fractional imaging).  Fiducial markers help increase the accuracy of these high-dose treatments, and automated techniques for marker identification open up new avenues of treatment guidance and monitoring.  Related to these trends, this work presents a fully automated workflow for simultaneously achieving two goals: (1) creating high quality templates of fiducial marker clusters, and (2) accurately identifying fiducial marker clusters in planar images.  It should be noted that, although marker tracking and template production are coupled in the presented routine, marker tracking can still be performed later, independent from template production.  Computationally intensive parts of the routine can be executed in advance by preparing templates beforehand using data from a prior scan.  Because of this structure, highly accurate real-time tracking could be accomplished with only a relatively inexpensive cross-correlation calculation.

Templates produced using the current technique have the potential to be particularly useful for intra-fractional monitoring during arc therapy techniques.  One common approach to real-time 3D tracking of markers is to use orthogonal kV/MV images acquired during treatment.\cite{Wiersma:2008kc, Liu:2008hr}  With unique templates being produced for all gantry angles, orthogonally tracked 2D positions would be able to pinpoint the target in 3D.  It should be noted that kV images examined in this work were acquired without an active MV beam, which has been known to degrade kV image quality.\cite{Fledelius:2014gq}  It would be worth investigating the performance of this technique in kV images that were acquired while a treatment beam was active, and in MV images acquired by a portal imaging device.  The cropping technique used in this work demonstrated an ability to maintain tracking in many instances where only a portion of the cluster was visible by the imager.  Such a technique could prove to be useful for target tracking in portal images of beams that are slightly off target or have been modulated by a multi-leaf collimator.

Regarding accuracy, the majority of errors observed in the phantom study were less than 100 $\mu$m.  Maximum errors for the 0D and 1D scans were smaller than the width of a single pixel, and the maximum error for the 3D scan was smaller than the width of two pixels.  By demonstrating sub-pixel accuracy, we have supported our choice to use the centroid of 9 pixels with the highest cross-correlation scores to determine the position of the cluster.  Furthermore, this shows that, moving forward, template matching techniques should be implemented in ways that allow for sub-pixel precision.  Had the current work been implemented based on single pixel position measurements, possible positions would have been limited to a grid with 259 $\mu$m between points, and errors would have been significantly larger as a result.

As for accuracy in patient studies, the relationship observed between FAHM and inferred absolute error offers a new metric for monitoring accuracy in cases where the ground truth is unknown.  In this work, 14 out of 15 patients had an average inferred absolute error smaller than the width of a single pixel, which is especially impressive when one considers how imprecise human measurements are known to be.  Results from the phantom study showed that distributions of inter- and intra-observer differences were wider than error distributions obtained by the automated tracking routine.  Previously, Harris et al considered fiducial marker positions as determined by three observers, and they found that intra-observer and inter-observer errors were on the order of 1.33 and 1.4 pixels, respectively.\cite{Harris:2006bg}  Regardless, manual tracking is not an option for intra-fractional marker monitoring.  As such, using FAHM to monitor the sharpness of cross-correlation peaks offers an intuitive approach for monitoring the accuracy of real-time template matching techniques.

The values observed for global tracking rate highlight the importance of the local tracking window.  An ideal rate of 100\% would indicate that the performance of the technique was not reliant on restricting the search to the local window.  Otherwise, the size of the local window becomes increasingly important as imaging frequency decreases.  In this work, a 21$\times$21 local window was small enough to maintain accurate tracking for all images, yet large enough to observe the maximum inter-frame shifts observed: 6.72 and 6.91 pixels in lateral and superior-inferior directions, respectively.  If imaging frequency were reduced, these maximum inter-frame shift values would increase, which could require an increase in the size of the local window.  As such, tracking techniques should strive to produce maxima that maintain their maximum status beyond the local window.  High global tracking rates are dependent on the target's appearance being distinct in comparison to other aspects of the projection image.  Many parts of a patient's anatomy can easily be confused for a single marker, especially when the marker is spherical or cylindrical.  Detection of false markers is a widespread issue when tracking markers individually, which calls for the use of elaborate methods of rejecting false markers.\cite{Pouliot:2001vh, Aubin:2003jx, Park:2009km, Fledelius:2011gb}  Our approach of seeking the entire cluster as a single entity produces templates that are much more distinct.

Compared to other recent template-based tracking techniques, our method compares favorably.  A method by Poulsen et al had a 99.9\% success rate and required that the user manually select certain projection images from the CBCT scan that featured large angular separation between them, good marker contrast, and a uniform background.\cite{Poulsen:2011dq}  An approach by Fledelius et al that required markers to be cylindrical in shape and used 3D constellation models for each individual marker to aid false marker rejection saw tracking success rates in the liver of 99.9\% and 99.8\%.\cite{Fledelius:2011gb, Fledelius:2014gq}  An approach by Regmi et al produced templates that were based on simulation CT scans with poor resolution (1.25 mm and 2.5 mm slice spacing), required manual selection of a volume-of-interest, and required manual corrections for variations in inter-marker spacing saw mean success rates of 100\%, 99.1\%, and 100\% for three tumor sites in CBCT images.\cite{Regmi:2014eb}  Our method offered greater simplicity, required no manual steps, did not require 3D constellation models for individual markers, and had a 100\% success rate.

The method introduced in this work also compares favorably to automated tracking methods that are not template-based.  Wan et al developed an iterative automated technique based on dynamic programming that tracks arbitrarily shaped coil markers in fluoroscopic images.\cite{Wan:2014jq, Wan:2016kv}  However, their algorithm minimizes a cost function based on the displacements of local minima observed for a set of multiple images; as coil markers can have many local minima on each marker, positions determined by their algorithm tend to jump around the marker, resulting in limited accuracy and precision (1.5$\pm$0.8 mm).\cite{Wan:2016kv}  Still, the remarkable speed of their approach (1-2 seconds for a set of CBCT projection images) could be useful for expediting the initial steps of our technique.  The Wan et al approach could be used to provide an initial rough tracking of markers, allowing for FME image stabilization before the first crude reconstruction of the marker cluster.  Stabilization prior to the first reconstruction would reduce the number of iterations necessary to reach a high quality template.

The current version of the algorithm has not yet been optimized for speed.  Beginning with raw projection images, automated target tracking and dynamic template production for a single CBCT scan can be performed in 30 minutes to an hour on a desktop personal computer (64-bit Windows 7, Intel\textsuperscript{\textregistered} Core\texttrademark$ $ i7-6700 CPU, 3.40 GHz, 16 GB RAM).  If one sought to use this technique simply to obtain target motion data, this speed is already more than sufficient and could be executed on a computer outside of the clinical workflow.  In order to immediately prepare templates for tracking targets in intra-fractional kV images, the speed of the algorithm would need to be improved.  However, if templates are prepared ahead of time, intra-fractional tracking is feasible.

Some parameters in the current work were reached empirically.  While these parameters would remain stable for a given anatomic site and imaging technique, they may need to be adjusted to accommodate other scenarios.  For projection images with larger amounts of noise, the size of filtering neighborhoods used in the initial image processing step may need to be increased.  For other marker materials, the density threshold used to clean up the reconstructed cluster volume may need to be adjusted.  Future work will test the current technique for other treatment sites to evaluate its performance under varying clinical conditions.

\section{Conclusions}
\label{Conclusions}

A novel, automated method for producing high quality, dynamic templates of fiducial marker clusters from a single CBCT scan has been introduced.  This method provides motion tracking data for planar imaging, and demonstrated a 100\% success rate for all fully visible marker clusters.  By cropping templates, tracking was also successful in many instances when portions of the cluster fell outside of the imager's field-of-view.  Initial results have shown that this method is remarkably accurate and robust in the face of many challenges commonly seen in fiducial marker tracking (e.g., radio-opaque non-marker objects, marker migration, deformation, rotation).  While only select patients saw better results when dynamic templates were used instead of static templates, the use of dynamic templates did not have a detrimental effect on tracking for other patients.  In addition to automatically providing clinically valuable motion data, templates produced using this method could potentially improve the reliability of real-time target tracking with intra-fractional kV imaging.

\section*{Acknowledgments}

This work was funded in part by the National Institutes of Health under award number K12CA086913, the University of Colorado Cancer Center/ACS IRG \#57-001-53 from the American Cancer Society, the Boettcher Foundation, and Varian Medical Systems.  These funding sources had no involvement in the study design; in the collection, analysis and interpretation of data; in the writing of the manuscript; or in the decision to submit the manuscript for publication.  The authors would also like to thank Dr. Tracey Schefter, Dr. Arya Amini, and Dr. Karyn A. Goodman for their support.  

\section*{Conflict of Interest Disclosure}

A provisional patent application has been submitted for the technique described in this work on behalf of the authors and the University of Colorado.


\begin{thebibliography}{10}

\bibitem[1]{Jaffray:2002tj}
D. A. Jaffray, J. H. Siewerdsen, J. W. Wong, and A. A. Martinez, ``{Flat-panel cone-beam computed tomography for image-guided radiation therapy},'' Int. J. Radiat. Oncol., Biol., Phys. \textbf{53}(5), 1337--1349  (2002).

\bibitem[2]{Moseley:2007dc}
D. J. Moseley, E. A. White, K. L. Wiltshire, T. Rosewall, M. B. Sharpe, J. H. Siewerdsen, J. Bissonnette, M. Gospodarowicz, P. Warde, C. N. Catton, and D. A. Jaffray, ``{Comparison of localization performance with implanted fiducial markers and cone-beam computed tomography for on-line image-guided radiotherapy of the prostate},'' Int. J. Radiat. Oncol., Biol., Phys. \textbf{67}(3), 942--953  (2007).

\bibitem[3]{Seppenwoolde:2002um}
Y. Seppenwoolde, H. Shirato, K. Kitamura, S. Shimizu, M. van Herk, J. V. Lebesque, and K. Miyasaka, ``{Precise and real-time measurement of 3D tumor motion in lung due to breathing and heartbeat, measured during radiotherapy},'' Int. J. Radiat. Oncol., Biol., Phys. \textbf{53}(4), 822--834  (2002).

\bibitem[4]{Berbeco:2005fd}
R. I. Berbeco, T. Neicu, and E. Rietzel, ``{A technique for respiratory-gated radiotherapy treatment verification with an EPID in cine mode},'' Phys. Med. Biol. \textbf{50}, 3669--3679  (2005).

\bibitem[5]{Poulsen:2008hs}
P. R. Poulsen, B. Cho, B. and P. J. Keall, ``{A method to estimate mean position, motion magnitude, motion correlation, and trajectory of a tumor from cone-beam CT projections for image-guided radiotherapy},'' Int. J. Radiat. Oncol., Biol., Phys. \textbf{72}(5), 1587--1596  (2008).

\bibitem[6]{Lam:1993uf}
K. L. Lam, R. K. Ten Haken, D. L. McShan, and A. F. Thornton, ``{Automated determination of patient setup errors in radiation therapy using spherical radio-opaque markers},'' Med. Phys. \textbf{20}(4), 1145--1152  (1993).

\bibitem[7]{Lewis:1995ue}
J. P. Lewis, ``{Fast normalized cross-correlation},'' Vision Interface \textbf{10}(1), 120--123  (1995).

\bibitem[8]{Balter:1995hy}
J. M. Balter, K. L. Lam, H. M. Sandler, J. F. Littles, R. L. Bree, and R. K. Ten Haken, ``{Automated localization of the prostate at the time of treatment using implanted radiopaque markers: technical feasibility},'' Int. J. Radiat. Oncol., Biol., Phys. \textbf{33}(5), 1281--1286  (1995).

\bibitem[9]{Harris:2006bg}
E. J. Harris, H. A. McNair, and P. M. Evans, ``{Feasibility of fully automated detection of fiducial markers implanted into the prostate using electronic portal imaging: a comparison of methods},'' Int. J. Radiat. Oncol., Biol., Phys. \textbf{66}(4), 1263--1270  (2006).

\bibitem[10]{Tang:2007jv}
X. Tang, G. C. Sharp, and S. B. Jiang, ``{Fluoroscopic tracking of multiple implanted fiducial markers using multiple object tracking},'' Phys. Med. Biol. \textbf{52}(14), 4081--4098  (2007).

\bibitem[11]{Cho:2009bj}
B. Cho, P. R. Poulsen, A. Sloutsky, A. Sawant, and P. J. Keall, ``{First demonstration of combined kV/MV image-guided real-time dynamic multileaf-collimator target tracking},'' Int. J. Radiat. Oncol., Biol., Phys. \textbf{74}(3), 859--867  (2009).

\bibitem[12]{Poulsen:2011dq}
P. R. Poulsen, W. Fledelius, P. J. Keall, E. Weiss, and J. Lu, ``{A method for robust segmentation of arbitrarily shaped radiopaque structures in cone-beam CT projections},'' Med. Phys. \textbf{38}(4), 2151--2156  (2011).

\bibitem[13]{Fledelius:2011gb}
W. Fledelius, E. Worm, U. V. Elstr{\o}m, J. B. Petersen, C. Grau, M. H{\o}yer, and P. R. Poulsen, ``{Robust automatic segmentation of multiple implanted cylindrical gold fiducial markers in cone-beam CT projections},'' Med. Phys. \textbf{38}(12), 6351--6361  (2011).

\bibitem[14]{Regmi:2014eb}
R. Regmi, D. M. Lovelock, M. Hunt, P. Zhang, H. Pham, J. Xiong, E. D. Yorke, K. A. Goodman, A. Rimner, H. Mostafavi, and G. S. Mageras, ``{Automatic tracking of arbitrarily shaped implanted markers in kilovoltage projection images: A feasibility study},'' Med. Phys. \textbf{41}(7), 071906 (12pp.)  (2014).

\bibitem[15]{Fledelius:2014gq}
W. Fledelius, E. Worm, M. H{\o}yer, C. Grau, and P. R. Poulsen, ``{Real-time segmentation of multiple implanted cylindrical liver markers in kilovoltage and megavoltage x-ray images},'' Phys. Med. Biol. \textbf{59}(11), 2787--2800  (2014).

\bibitem[16]{Jones:2015dj}
B. L. Jones, D. Westerly, and M. Miften, ``{Calculating tumor trajectory and dose-of-the-day using cone-beam CT projections},'' Med. Phys. \textbf{42}(2), 694--702  (2015).

\bibitem[17]{Jones:2015jz}
B. L. Jones, T. Schefter, and M. Miften, ``{Adaptive motion mapping in pancreatic SBRT patients using Fourier transforms},'' Radiother. Oncol. \textbf{115}(2), 217--222  (2015).

\bibitem[18]{Savitzky:1964bn}
A. Savitzky and M. J. E. Golay, ``{Smoothing and differentiation of data by simplified least squares procedures}," Anal. Chem. \textbf{36}(8), 1627--1639  (1964).

\bibitem[19]{Wiener:1949}
N. Wiener, \emph{Extrapolation, Interpolation, and Smoothing of Stationary Time Series}, (Wiley, New York, 1949).

\bibitem[20]{Wiersma:2008kc}
R. D. Wiersma, W. Mao, and L. Xing, ``{Combined kV and MV imaging for real-time tracking of implanted fiducial markers},'' Med. Phys. \textbf{35}(4), 1191--1198  (2008).

\bibitem[21]{Liu:2008hr}
W. Liu, R. D. Wiersma, W. Mao, G. Luxton, and L. Xing, ``{Real-time 3D internal marker tracking during arc radiotherapy by the use of combined MV-kV imaging},'' Phys. Med. Biol. \textbf{53}(24), 7197--7213  (2008).

\bibitem[22]{Park:2009km}
S. J. Park, D. Ionascu, F. Hacker, and H. Mamon, ``{Automatic marker detection and 3D position reconstruction using cine EPID images for SBRT verification},'' Med. Phys. \textbf{36}(10), 4536--4546  (2009).

\bibitem[23]{Pouliot:2001vh}
S. Pouliot, A. Zaccarin, and D. Laurendeau, ``{Automatic detection of three radio-opaque markers for prostate targeting using EPID during external radiation therapy},'' International Conference on Image Processing \textbf{2}, 857--860  (2001).

\bibitem[24]{Aubin:2003jx}
S. Aubin, L. Beaulieu, S. Pouliot, J. Pouliot, and R. Roy, ``{Robustness and precision of an automatic marker detection algorithm for online prostate daily targeting using a standard V-EPID},'' Med. Phys. \textbf{30}(7), 1825--1832  (2003).

\bibitem[25]{Wan:2016kv}
H. Wan, J. Bertholet, J. Ge, P. Poulsen, and P. Parikh, ``{Automated patient setup and gating using cone beam computed tomography projections},'' Phys. Med. Biol. \textbf{61}(6), 2552--2561  (2016).

\bibitem[26]{Wan:2014jq}
H. Wan, J. Ge, and P. Parikh, ``{Using dynamic programming to improve fiducial marker localization}," Phys. Med. Biol. \textbf{59}, 1935--1946  (2014).

\end{thebibliography}

\end{document}